\documentclass[a4paper,fleqn,usenatbib]{mnras}

\usepackage{newtxtext,newtxmath}

\usepackage[T1]{fontenc}
\usepackage{ae,aecompl}
\usepackage{tablefootnote}

\usepackage{graphicx}	
\usepackage{amsmath}	
\usepackage{longtable}
\usepackage{pdflscape}

\title[GX339-4, A Minimal Timescale]{A Minimal Time Scale for the Spectral States of GX 339-4}

\author[E. Sonbas et al.]{
E. Sonbas,$^{1,2}$\thanks{E-mail: edasonbas@gmail.com}
K. Mohamed,$^{2,3}$
K. S. Dhuga$^{2}$
A. Tuncer$^{4}$
and E. G\"o\u{g}\"u\c{s}$^{5}$
\\
$^{1}$Adiyaman University, Department of Physics, 02040 Adiyaman, Turkey\\
$^{2}$Department of Physics, The George Washington University, Washington, DC 20052, USA\\
$^{3}$Department of Physics, Sohag University, Egypt \\
$^{4}$Istanbul University Science Faculty, Department of Astronomy and Space Sciences, 34119, University-Istanbul, Turkey\\
$^{5}$Sabanc\i~University, Orhanl\i~- Tuzla, Istanbul 34956, Turkey
}

\date{Accepted 2020 September 17; Received  2020 September 15; in original form 2020 May 31}

\pubyear{2020}

\begin{document}
\label{firstpage}
\pagerange{\pageref{firstpage}--\pageref{lastpage}}
\maketitle
\begin{abstract}
\noindent
Black hole transients are known to undergo spectral transitions that form q-shaped tracks on a hardness intensity diagram. In this work, we use archival RXTE data to extract a characteristic minimal timescale for the spectral states in GX 339-4 for the 2002-3, and 2010 outbursts. We use the extracted timescale to construct an intensity variability diagram for each outburst. This new diagram is comparable to the traditional hardness intensity diagram and offers the potential for probing the underlying dynamics associated with the evolution of the relevant emission regions in black hole transients. We confirm this possibility by connecting the minimal timescale with the inner disk radius, $R_{in}$ (estimated from spectral fits), and demonstrate a positive correlation between these variables as the system evolves through its spectral transitions. Furthermore, we probe the relation between the minimal timescale and the break frequencies extracted from the power spectral densities. Lastly, we examine a possible link between the extracted timescale and a traditional measure of variability i.e., the root mean square, determined directly from the power spectra.
\end{abstract}
\section{Introduction}

\noindent
Studies show that black hole (BH) transients undergo transitions among various spectral states. These transitions and the states are usually depicted on a hardness intensity diagram (HID), and are extensively discussed in the literature \citep{2006ARA&A..44, 2010LNP...794...53B}. The HID is essentially a schematic that presumably encapsulates the degree of coupling between the accretion rate and the relative configuration of the accretion disk and is typically used to track the transitions between the spectral states. An example of the various spectral states and the regions they occupy on a HID is shown by \citet{2005A&A...440..207B} for the BH transient GX 339-4. There is universal agreement for the two most prominent states which are the Low/Hard (LHS) and the High/Soft states (HSS). Other states such as the Hard/Intermediate state (HIMS), and the Soft/Intermediate state (SIMS) have also been identified \citep{2007A&ARv..15....1D, 2005Ap&SS.300..107H,  2010MNRAS.403...61D}. The quiescent state, which often gets linked together with the LHS rounds out the categories.\\
\\
\noindent
Other spectral states have been proposed: one such set \citep{2006ARA&A..44} contains three spectral states: hard, thermal and the steep power law (SPL). Based on spectral and timing properties, the hard and the thermal states are readily identified with the LHS and HSS respectively. However, the SPL state is not so easily characterized; it apparently encompasses or overlaps the two intermediate states i.e., HIMS and SIMS. This is potentially a source of confusion and uncertainty especially when comparing spectral properties of a sample of BH systems based solely on a luminosity scale. Nonetheless, the HID continues to be used as a comparative tool for tracking state transitions, as well as studying the underlying coupling of the accretion rate, the disk, and the surrounding corona.\\
\\
Besides the accretion rate itself, the importance of the disk and the corona, cannot be underestimated as their configurations and dynamic interplay are thought to be the origin of the main emission components that determine the majority of the observed spectral and temporal properties of BH binaries. It is postulated that as a particular source undergoes a state transition, the extent of the disk is dynamically coupled to the particular state of the source. While qualitatively plausible, there are only a few robust instances of evidence \citep{2014MNRAS.442.1767P, 2009ApJ...707L..87T} to support the actual extent to which the disk size changes and its dynamic relation to either the size of the corona or the spectral state of the source. Indeed some studies seem to suggest that the disk may not be truncated in the low/hard (LHS) state \citep{2009MNRAS.395L..52R, 2010ApJ...709..358R, 2013ApJ...769...16R}. Clearly, in order to quantitatively determine the extent and the evolution of the disk/corona configuration as sources undergo transitions, much more rigorous and systematic temporal and spectral information, that goes beyond that available via the simple HID, is required over a large luminosity range, covering all accessible spectral states.\\
\\
\noindent
Another parameter (in addition to the hardness ratio in the HID) that has proven useful is the root mean square (RMS), a traditional measure of variability which is determined by integrating the power spectral density (PSD) in a specific frequency range \citep{1983ApJ...266..160L, 1988SSRv...46..273L, 1989ARA&A..27..517V, 1990A&A...230..103B, 1992ApJ...391L..21M}. In this work, we propose the use of a different parameter, one based on a temporal property, to enable the monitoring of spectral states and their transitions. The scientific motivation is straightforward; we seek a minimal timescale (MTS) that can be estimated directly from lightcurves and be used to track the transitions in a way that is both complementary and not too dissimilar from the function of the hardness ratio in HIDs. We interpret the proposed timescale to imply the smallest temporal feature in the lightcurve that is consistent with a fluctuation above the Poisson noise level. In the frequency domain, this would be equivalent to the highest frequency component in the signal at or just above the noise threshold. Assuming that the (intrinsic) signal-noise threshold is different for different states, the proposed timescale implies a temporal tag for each state and associated transitions, thus providing a tool that can be used to track their dynamic evolution. A robust method for extracting such a measure was extensively described by \citet{2013MNRAS.432..857M}, in which the authors used a technique based on wavelets to analyze lightcurves in the time domain for a sample of gamma-ray bursts and demonstrated the robust separation of white noise (background) from red noise (signal) at a level of a few milliseconds. Of course, characteristic timescales can also be accessed via power spectral densities i.e., PSDs in the frequency domain. In this study, we present the extraction of the MTS in the time domain as an alternative and compare our results with timescales extracted from the PSDs, and with other properties, such as the RMS.\\
\\
The layout of the paper is as follows: in Section 2, we describe the selection and extraction of the available RXTE data for GX 339-4; Section 3 contains the details of the computation of the MTS, the RMS, and their comparison, as well as, the utility of MTS in the construction of an intensity 'variability' diagram (IVD). For a comparison of the MTS with another timescale, we extract the break frequencies from the fits of power spectral densities. In addition, we describe the extraction of the inner disk radius, $R_{in}$, from spectral fits and probe its possible connection with MTS. The implications of these results are discussed in the same section. We conclude by summarizing our main findings in Section 4.
\section{Data Reduction and Methodology}
\noindent
We analyzed publicly available RXTE/PCA observations covering the 2002-3 and 2010 outbursts of GX 339–4, a source that has hundreds of RXTE observations (see Table 1) across its spectral states and transitions. We used the standard RXTE data analysis using the tools of HEASOFT V.6.24 to create lightcurves with a time resolution ranging from 200 $\mu$s to several ms. Where possible, we used high resolution Good-Xenon or Event data modes that cover the full energy band, otherwise we combined Single-Bit and Event data modes to cover the full energy band. The primary reason for using the full energy band was to obtain a measure of the background noise which would appear in the highest available channels. The high resolution lightcurves were created using standard screening criteria by applying time filtering with the criteria on elevation angle to be greater than 10$\degr$, and an offset from the source to be less than 0.02$\degr$.\\
\\
We extracted  MTS associated with the state transitions of GX339-4 by using a fast wavelet transformation to encode light curves into a wavelet representation (coefficient expansion) and then compute a statistical measure of the variance of wavelet coefficients over multiple time-scales, a technique we developed to extract temporal scales of gamma-ray bursts. Wavelet transformations have been shown to be a natural tool for multi-resolution analysis of non-stationary time-series \citep{Flandrin89, Flandrin92, Mallat89}. Wavelet analysis is similar to Fourier analysis in many respects but differs in that wavelet basis functions are well-localized, \emph{i.e.} have compact support, while Fourier basis functions are global. Compact support means that outside some finite range the amplitude of wavelet basis functions goes to zero or is otherwise negligibly small \citep{Percival00}. In principle, a wavelet expansion forms a faithful representation of the original data, in that the basis set is orthonormal and complete. In our case, the signal of interest of course is the measured X-ray lightcurve of GX339-4. It is fairly well established that the emission process most likely involves more than one mechanism. Moreover, we know that the accretion rate is unlikely to be constant over time especially since we know that the GX339-4 undergoes spectral transitions involving a large variation in intensity. Under these conditions, we can expect a non-stationary component in the underlying emission process.\\ 
\\
The main utility of wavelets is that they can be stretched and compressed in time i.e., altering the frequency content of the wavelet with each operation. This is achieved by adjusting the so-called scale factor; the larger the scale the larger the spread of the wavelet in time. Conversely, the smaller the scale, the more compressed the wavelet in time, corresponding to a higher frequency wavepacket. The freedom to adjust the scale of the wavelet enables one to search for time structures in a given lightcurve that matches the scaled wavelet. All that is required is a scan of the entire lightcurve in a systematic way, at different scales. The scanning is achieved by the so-called translation factor which simply moves the location of the wavelet function from time zero by incremental time steps until the entire duration of the lightcurve has been covered. The result of each step is coded in terms of coefficients (essentially the product of the wavelet function and the lightcurve). These coefficients (in terms of the scale and translation factors) provide a measure of the power in the signal. Clearly one only obtains finite power where there is significant overlap (convolution) of the wavelet function and some structure at the right temporal scale in the lightcurve. If a significant overlap is not found the coefficients tend to be small reflecting little power at that scale. The process of scanning the lightcurve for every scale is repeated and the appropriate coefficients extracted.\\
\\
Although many wavelets are available, we used one of the simplest functions i.e., the Haar wavelet to represent the observed light curves in terms of detail ($d_{jk}$) coefficients. The variance of these coefficients is used to construct a logscale diagram i.e., a plot of log of variance ($\rho_j$) vs. frequency (octaves).
\begin {equation}
\rho_j = \frac{1}{n_j}\sum_{k=0}^{n_j -1} |d_{jk}|^2
\end {equation}
\begin {equation}
\log_2(\rho_j) = \lambda j + constant
\end {equation}
This plot is then used to identify the scaling (signal) and noise (evenly distributed power as function of frequency) regions and extract the variability time scale associated with a transition. White-noise (background) processes appear in logscale diagrams as flat regions while non-stationary (weakly or otherwise) processes, including the signal/red noise of interest, appear as sloped regions with slope $\lambda$ (see equation (2)). The transition between these regions occurs at some characteristic time scale referred to as the minimum variability 'time' scale (MTS); A regression method is utilized to determine the intersection of these two regions and hence the MTS, $\it{j}$, in octaves (see \citet{2013MNRAS.432..857M}). An example of a logscale diagram for a particular observation (Obs. ID: 70110-01-08-00: 2002-3 outburst) is shown in Figure 1 (for a light curve with a time binning of 1 ms) along with the regression fit used in extracting the intersection point. The octave scale is readily converted to a real time scale by using the binning time of the light curve data (i.e., the MTS (s) $\sim$ 2$^j$ x timebin). In few cases, mostly the high-soft and soft-intermediate states where the PSDs are irregularly shaped and/or are dominated by low signal-to-noise ratios (SNR), the logscale diagrams show considerable scatter and large error bars making regression fits overly sensitive to the fit region. An example of such a logscale diagram is shown in Figure 2. In such cases, we opted for a simpler method of extracting the MTS; the lighcurves were divided into several sections and we constructed the logscale diagram for each section. These separate diagrams are depicted by different colors in Figure 2. As can be seen, the different diagrams indicate different levels of background noise in the flat region. In order to extract the MTS, we proceeded as follows: using the minimum and maximum deviation of the log(variance) (in the flat region), we determined the mean background level. Then we scanned the logscale diagram toward the sloped region and determined the first 3$\sigma$-point of deviation from the mean background level. That point we took to be the intersection, i.e., the octave $\it{j}$, which then could be converted to MTS. We also summed the logscale diagrams in order to confirm the extracted $\it{j}$ and, importantly, this allowed us to examine the general trend of the variance in the intersection region and make sure that the selected (octave) point was a part of the increasing variance as a function of the octave scale instead of some outlier that happened to be 3$\sigma$ from the mean of the noise level. The typical error bar for the extracted $\it{j}$ was the order of $\pm$ 0.5 – 1.0 (octave) for both the regression and the 3$\sigma$ methods. For additional details, such as the simulation method used to estimate the effects of background subtraction and the reconstruction of the signal from the detailed coefficients, the reader is referred to the work of \citet{2013MNRAS.432..857M}.\\
\begin{figure}
\hspace{-1.1cm}
\centering      
\includegraphics[scale=0.60, angle = 0]{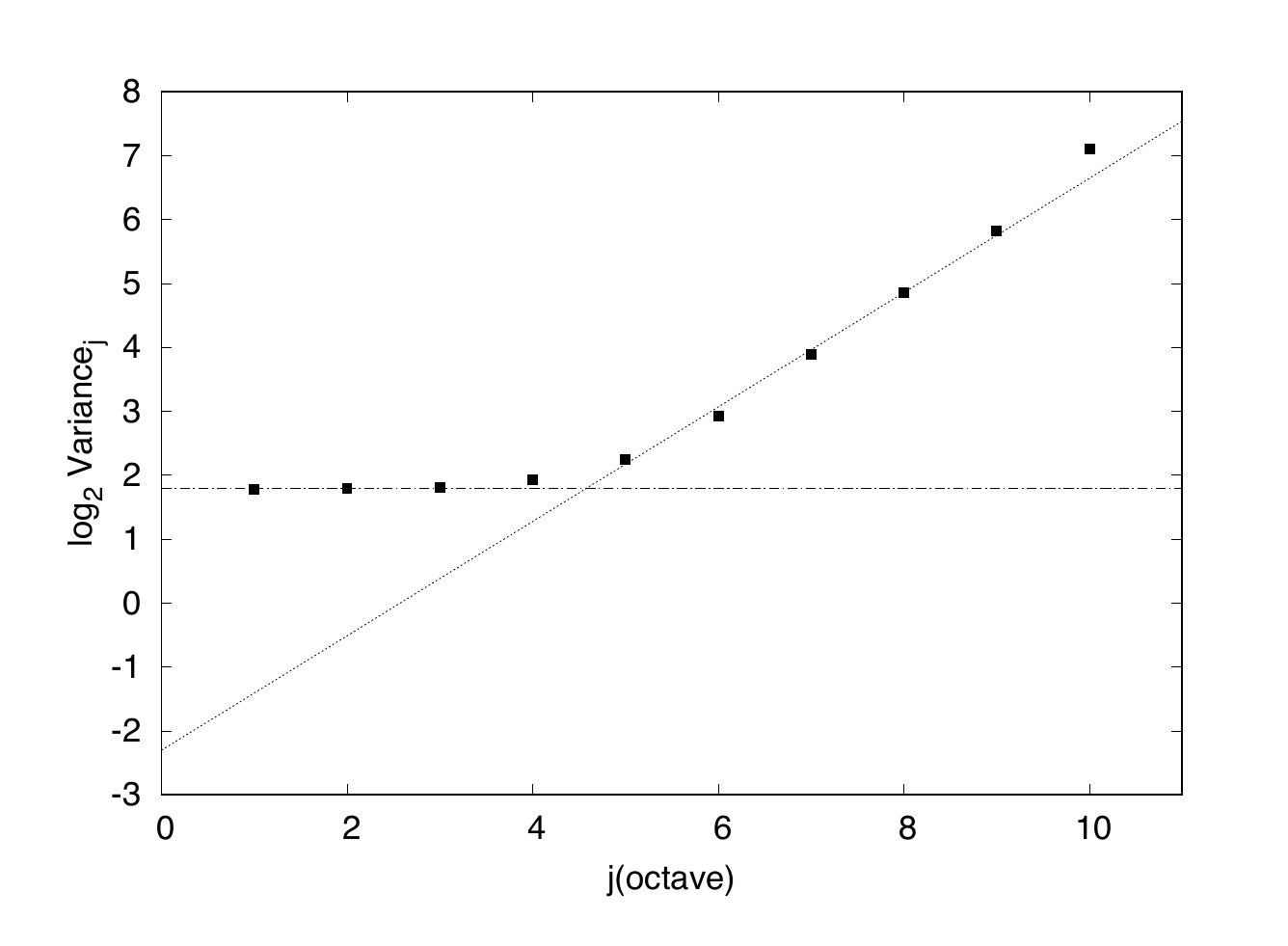}   
\caption{Log (variance) vs octave (j) for a particular observation (obsid:70110-01-08-00). The flat region indicates background noise, the sloped region is the signal of interest. The regression fit line, to obtain the interection, is also shown.}
    \label{fig1}
\end{figure}
\begin{figure}
\hspace{-1.1cm}
\centering      
\includegraphics[scale=0.60, angle = 0]{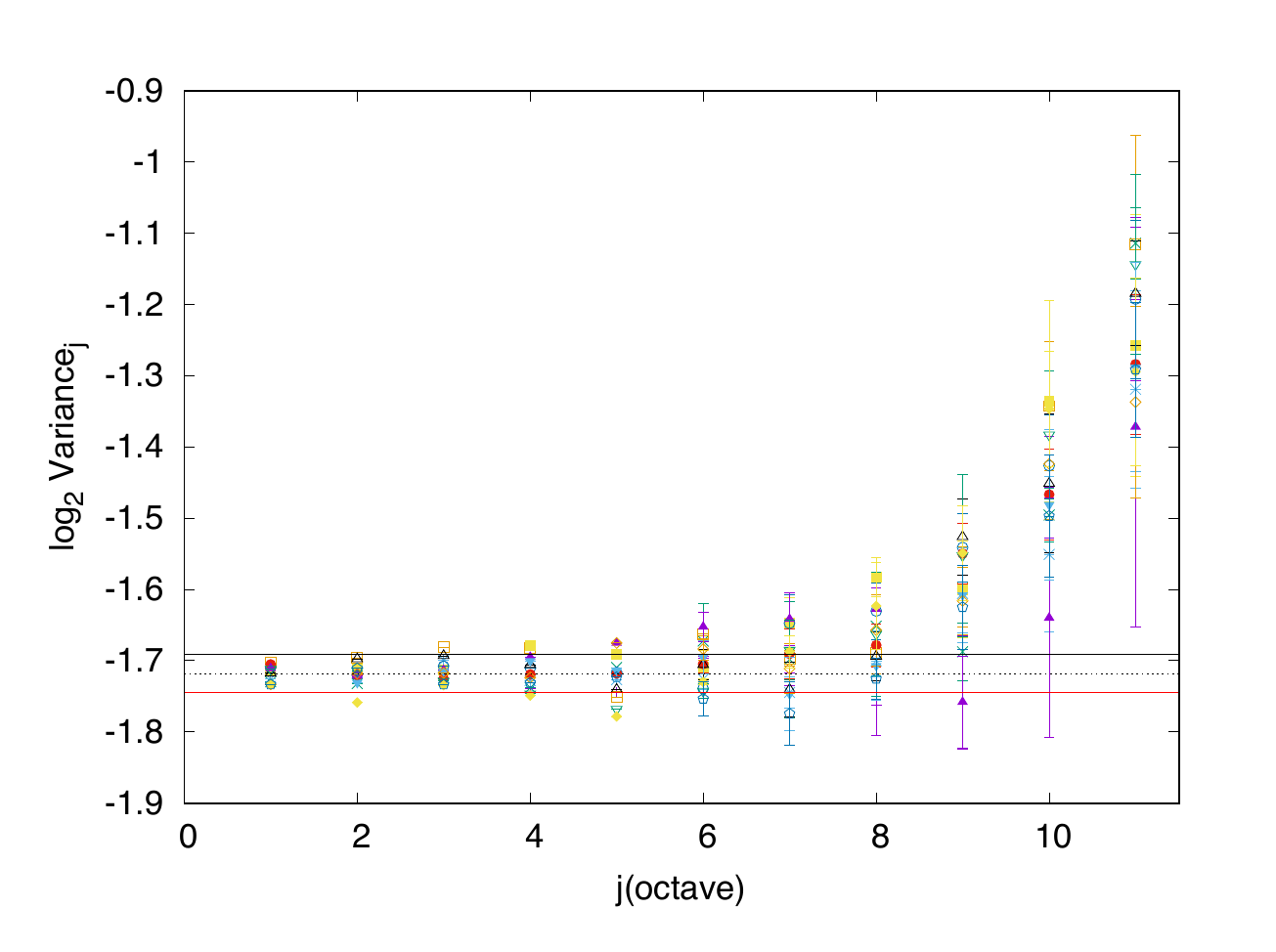}   
\caption{Log (variance) vs octave (j) for obsid:70110-01-79-00. The different colors indicate different levels of background noise. The minimum, mean, and maximum variation in noise region indicated by horizontal lines.}
    \label{fig2}
\end{figure}
\\
We normalized the PSDs following \citet{1989A&A...225...79H}, \citet{1990A&A...230..103B} , and \citet{1992ApJ...391L..21M}  taking into account the background renormalization and Poisson level subtraction. The total RMS was computed directly from the integration of the averaged PSDs in the frequency band [0.1-64 Hz]. Also, we computed the absolute rms variability, $\sigma(rms)$, which equals the RMS multiplied by the mean count rate of the source \citep{2001MNRAS.323L..26U}. The mean count rate for PCU2 and the hardness ratio for each observation were taken to be the same as \citet{2005A&A...440..207B} with a slight shift in the energy bands according to the last update of epoch 5 i.e., the count rates were averaged in the PCA channel range 8 - 49 (3.3 - 21.0 kev) and the hardness ratio was calculated using the channel ranges: 8 - 14 (3.3 - 6.1 keV) and 15 - 24 (6.1 - 10.3 keV) respectively. For the analysis, we used custom software under HEASOFT (v. 6.24) and IDL: binlc.pro (v. 1.25-2008) following \citet{2005A&A...440..207B} , \citet{2000MNRAS.318..361N}, and \citet{2003A&A...407.1039P}. The noise level was subtracted taking into account the dead time corrections, see \citet{1999ApJ...510..874N}, \citet{1995ApJ...449..930Z}, \citet{1996ApJ...469L..29Z}, and \citet{2000ApJ...530..875J}. The uncertainties in RMS and $\sigma$(rms) were computed using standard error propagation methods (\citet{1989ARA&A..27..517V}, and \citet{2004A&A...414.1091G}). 
\section{Results and Discussion}
\noindent
As noted in the introduction, the HID and the q-curve behavior of its parameters as BH binaries undergo transitions from the low/hard state to the thermally dominated soft state through a number of intermediate states (HIMS, SIMS etc) and back down to the low/hard state and finally quiescence, is well established. We present in Figures 3 and 4, the IVDs i.e., the intensity (count/rate) - MTS plots for GX 339-4 (2002-3 and 2010 outbursts). For comparison, we also constructed the HIDs (see Figure 5; 2010 outburst indicated in grey) for the same set of observations that were deployed for the IVDs. Despite the greater scatter in the MTS plots, the IVDs follow a similar q-like pattern to that observed in the traditional HID in that the MTS is sensitive to, and tracks, the the various spectral states of the source. The black and red colors (in the IVD) represent rise and decay parts of the LHS and HIMS transitions respectively. Specifically, the MTS tracks the low/hard (LHS; circles, black and red), high/soft (HSS; blue triangles), hard intermediate state (HIMS; squares, black and red), and the soft intermediate state (SIMS; pink crosses) for GX 339-4. Note that the HSS state is well separated from the LHS and HIMS states on the temporal scale, with the HSS, on average, corresponding to a significantly smaller time scale. The SIMS, on the other hand, correlates with a time scale intermediate to those for the LHS and HSS states. On the whole, not only does it appear that the separation of states is more pronounced in the 2010 outburst, the transitions too appear to be more orderly and smooth in comparison to the somewhat erratic features seen in the 2002-3 outburst.\\
\\
\begin{figure}
\hspace{-1.1cm}
\centering      
\includegraphics[scale=0.40, angle = 0]{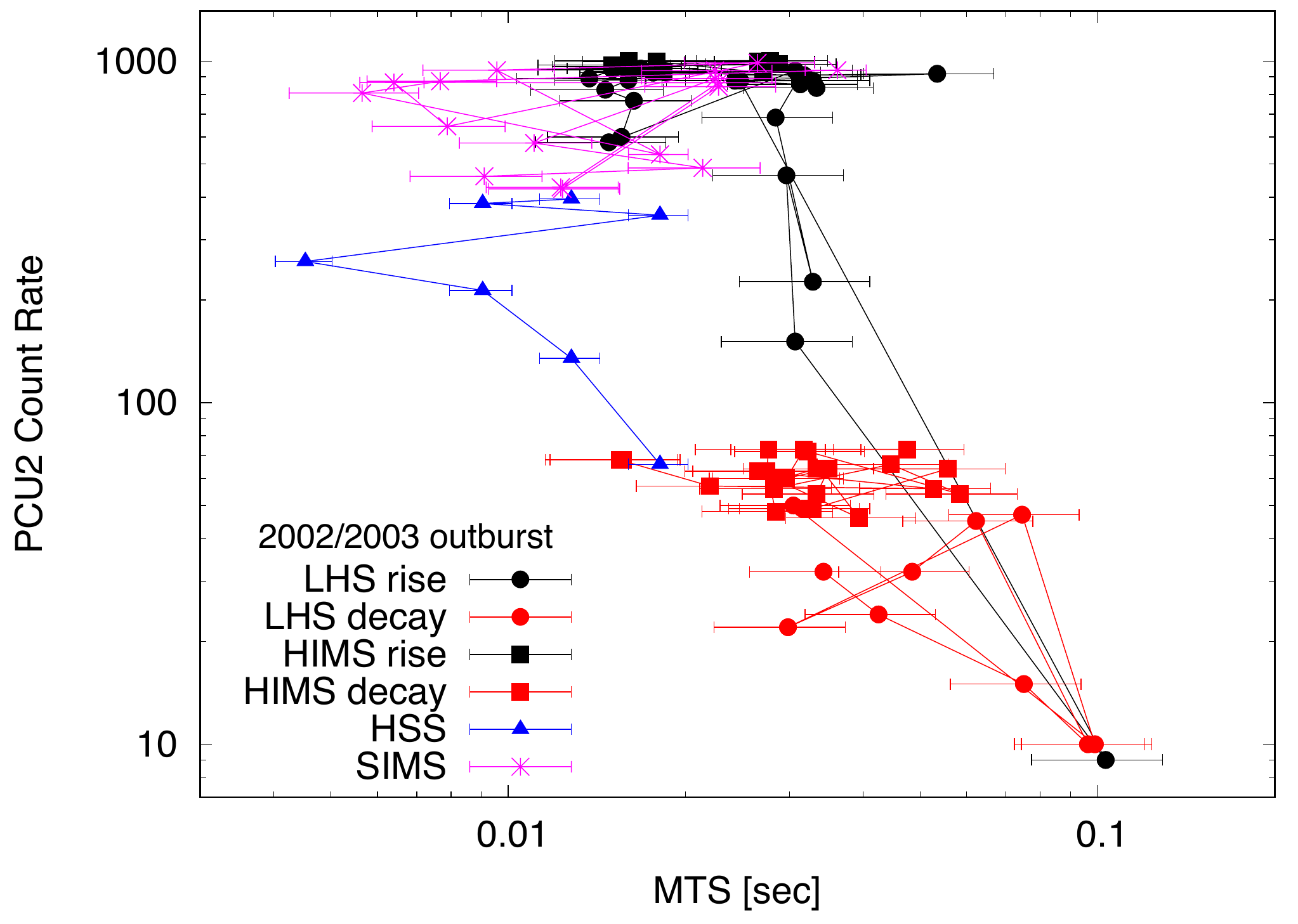}   
\caption{Intensity vs MTS for GX 339-4 for the 2002-3 outburst: The black and red colors represent rise and decay parts of the LHS and HIMS transitions respectively. The HSS state (blue triangles) is well separated from the LHS state (circles) and HIMS state (squares) on the temporal scale. The SIMS (pink crosses) appears to correspond to a time scale intermediate to those for the LHS and the HSS states.}
    \label{fig3}
\end{figure}
\begin{figure}
\hspace{-1.1cm}
\centering      
\includegraphics[scale=0.40, angle = 0]{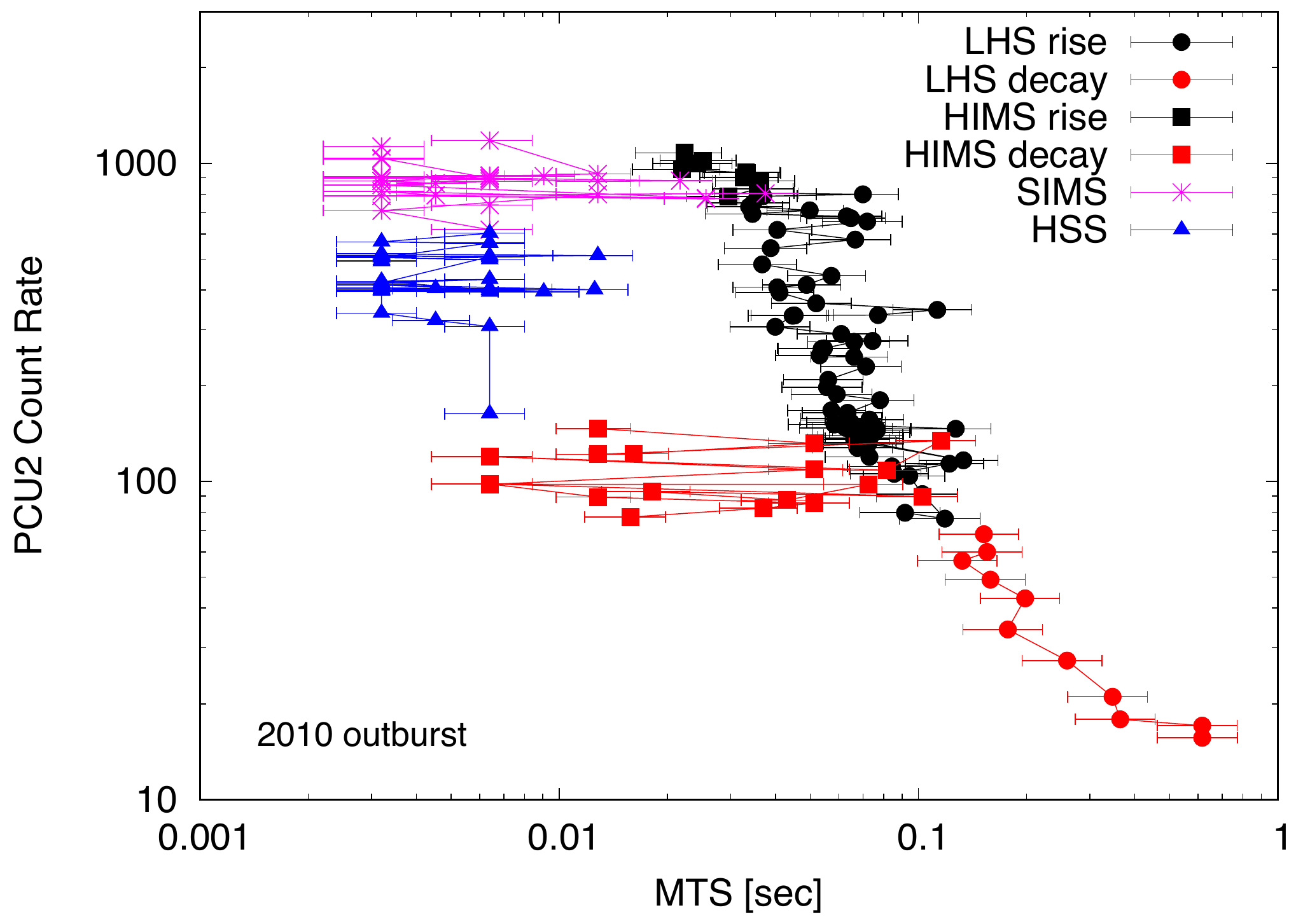}   
\caption{Intensity vs MTS for GX 339-4 for the 2010 outburst; The colors and the symbols are the same as in Figure 3. The q-shape is pronounced and the separation in states is clearly evident.}
    \label{fig3}
\end{figure}
\begin{figure}
\hspace{-1.1cm}
\centering      
\includegraphics[scale=0.65, angle = 0]{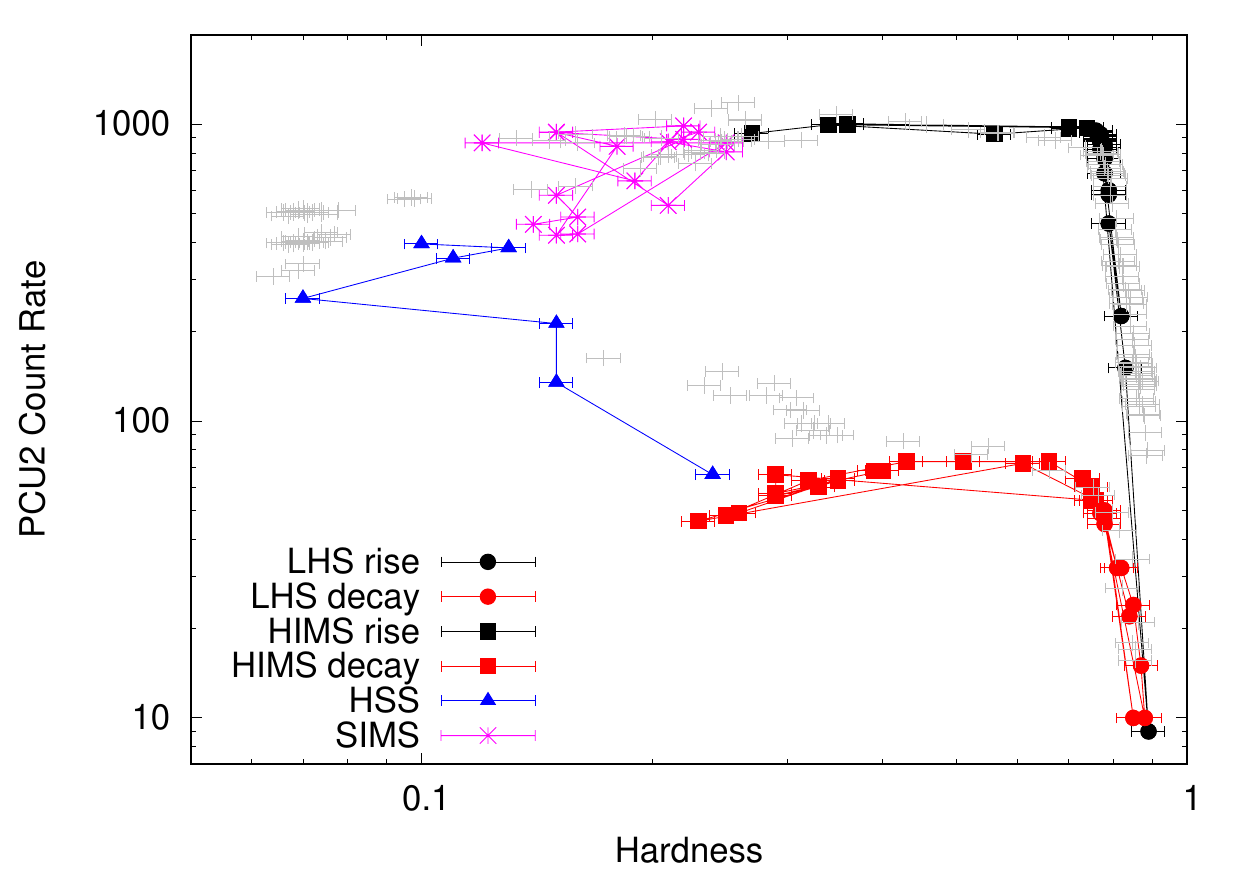}    
\caption{Intensity vs Hardness for GX 339-4 for the 2002-3 and 2010 (grey) outbursts. The similarity between Figures 3, 4, and 5 is evident although the scatter is clearly larger in the MTS plots.}
    \label{fig4}
\end{figure}
\\
A standard measure of the variability in lightcurves is the RMS. We separate this measure into its two most commonly used forms: $\sigma(rms)$ and fractional RMS. $\sigma(rms)$ is essentially a measure of the variance of a time series and is normally computed (in counts/s) directly from the lightcurve.  The fractional RMS, on the other hand, is normalized with respect to the count rate and is typically extracted from a suitably normalized PSD (and quoted as a percentage). The RMS, extracted in a well-defined frequency range, is useful in studying the variability as a function of flux. $\sigma(rms)$ is well known to exhibit a linear correlation with flux \citep{2001MNRAS.323L..26U} for the low-hard state (LHS).  \citet{2011MNRAS.410..679M} introduced the rms-intensity diagram (RID), which made it possible for them to probe the BH transitions without the aid of spectral information. Using the RID (their Figure 1, equivalent to our Figure 6), they were able distinguish several regions corresponding to various stages of spectral transition in GX339-4. In particular, the RID isolates the so-called hard-line transition corresponding to the low-hard state (LHS) where the linearity of the rms-flux relation is very clear. However, the relation becomes progressively more complex as the system executes an anticlockwise hysteresis loop. The $\sigma(rms)$-flux relation is observed for X-ray binaries and AGN, and is seen across multiple time scales suggesting an underlying intrinsic nature involving correlated processes. If MTS is somehow linked to a variability measure  then an interesting question arises as to how the two measures i.e., RMS and MTS are related?  The prime reason for exploring a potential connection between these two variables is simple in that RMS is a traditional measure of variability and we would like to determine whether MTS can serve as a proxy for RMS. We certainly do not expect a causal relation (i.e., correlation does not necessarily imply causation) but nonetheless it is possible that some correlation exists between the two that might be useful in exploring a hidden variable that both RMS and MTS independently depend on. We examine this in the following section.\\
\\
From a theoretical perspective, a possible model \citep{1997MNRAS.292..679L} that potentially explains the $\sigma (rms)$-flux relation assumes propagation of pulses/fluctuations of the accretion rate in the disk. These fluctuations are presumably thermally generated and/or viscosity driven, and therefore, the characteristic time MTS, can in principle, be thought of as a probe of the variability at the relevant time scale for this flow. To set the stage, we extracted the fractional RMS for all the spectral states available in the observations. We also computed $\sigma(rms)$. Our results of flux vs. $\sigma (rms)$ are shown in Figure 6: we make a couple of observations; the linear correlation for the LHS state (black and red circles representing the rise and the decay of the outburst respectively) is clearly seen, and secondly, as noted by \citet{2011MNRAS.410..679M}, the $\sigma (rms)$-flux correlation is somewhat more complicated for the HIMS, SIMS, and the HSS states. In Figure 7, we show the resulting fractional RMS - MTS plots for the 2002-3 (upper panel) and 2010 outbursts (lower panel). The respective best-fits are shown as solid lines (black; 2002-3, and grey; 2010). Although the dispersion is relatively large, fractional RMS and MTS exhibit a positive trend i.e., low RMS variability corresponds to short time scales and high RMS variability correlates with longer time scales. The slope of the best-fit line for 2002-3 outburst is 1.9 $\pm$ 0.2, and significantly shallower i.e., 0.59 $\pm$ 0.01, for the 2010 outburst. The correlation is relatively strong for the 2010 outburst (a Pearson correlation coefficient of 0.88; null probability $\sim$ 2 x$10^{-21}$) compared to the 2002-3 outburst (a coefficient of 0.53; null probability $\sim$ 6 x$10^{-7}$). We also notice that the plots shows a moderate separation (in time) between the LHS and the HSS states, with the LHS state exhibiting, on average, a larger time scale ($\sim$30 ms) than the HSS state ($\sim$ 10 ms). The difference in slope is mostly due to the fact that the MTS scale for the 2010 outburst has been extended (toward larger times) by the decay part of the low-hard state (full red circles; lower panel); the extension is almost a factor 4 compared to the decay part of the low-hard state for the 2002-3 outburst (upper panel). As the effect seems to be related to the decay part of the low-hard state this would suggest a possible connection with the corona, however, the nature of this connection, if there is one, is not clear at this stage. The observation id's, RMS variabilities, and the extracted MTS for the two outbursts are listed in Tables 1 and 2 respectively. In Figures 8 and 9, we show the plots of RMS and MTS vs Hardness for the 2002-3 and 2010 outbursts. The upper portions of the figures depict the RMS($\%$) and the lower portions, the MTS (sec). We note that the spectral states are essentially differentiated in the MTS-hardness plane in the same manner as in the RMS-hardness plane. This is more noticeable for the 2010 outburst (Figure 9) as compared to the 2002-3 outburst (Figure 8) where the effect is minor. Again, the greater dispersion in the MTS-hardness plane is notable. In particular the LHS state (filled black circles), depicting the hard portion of the spectrum, is separated in MTS from the HSS (and SIMS) (blue and lilac points, depicting the softer portion of the spectrum) by almost an order of magnitude (see Figures 3 and 4, where it is more apparent). In Figure 9, we do point to one distinction between the MTS and RMS: for hardness ratio $\sim$ 0.7, we notice that the MTS-hardness plot clearly seems to be differentiating betweeen the rising part of the low-hard state (black circles) and its decay part of the transition (red circles) much more cleanly than the RMS portion of the plot. \\
\begin{figure}
\centering      
\includegraphics[scale=0.60, angle = 0]{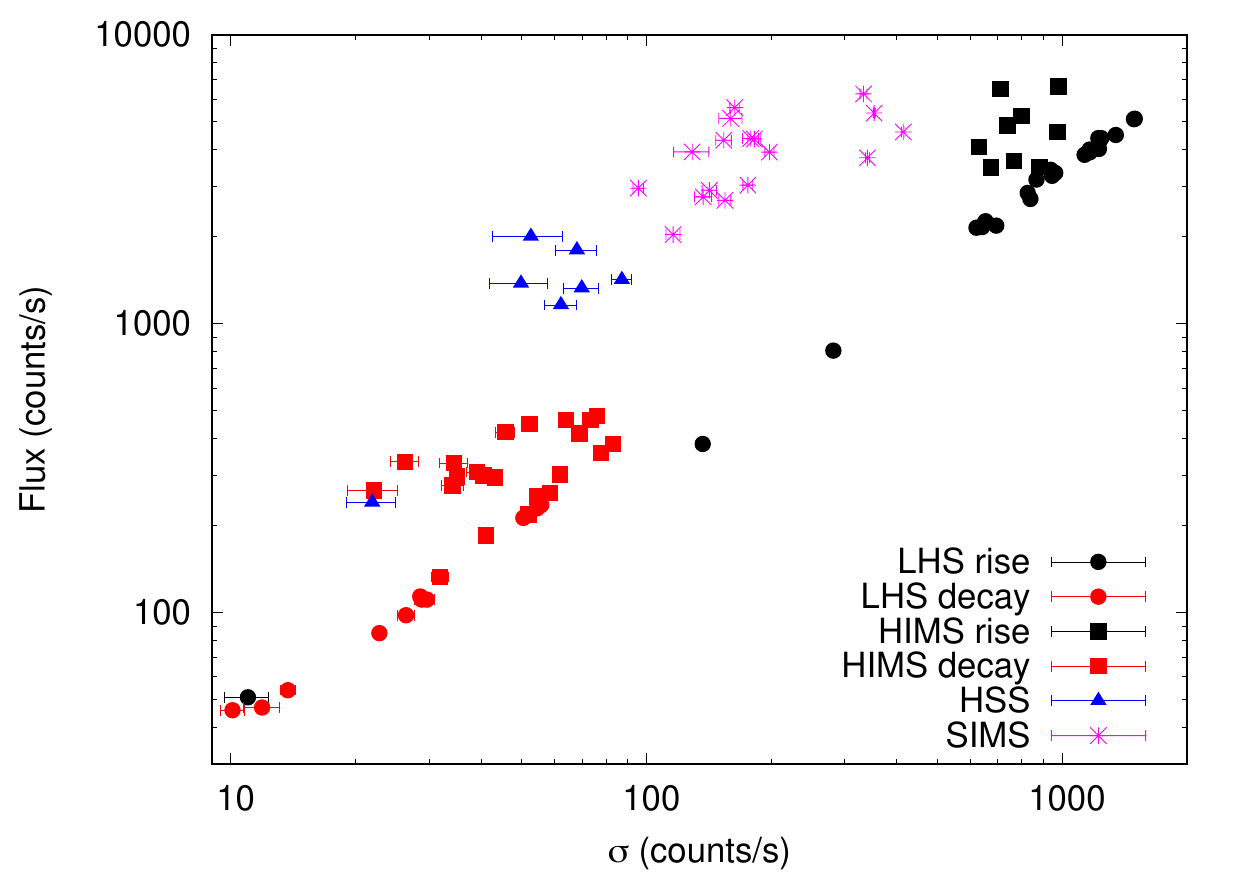}
\caption{Flux vs $\sigma$ (rms) for the spectral states in GX339-4 for the 2002-3 outburst. The linear portion is evident for the LHS state (black and red circles representing the outburst rise and decay respectively).}
    \label{fig6}
\end{figure}
\begin{figure}
\centering      
\includegraphics[scale=0.35, angle = 0]{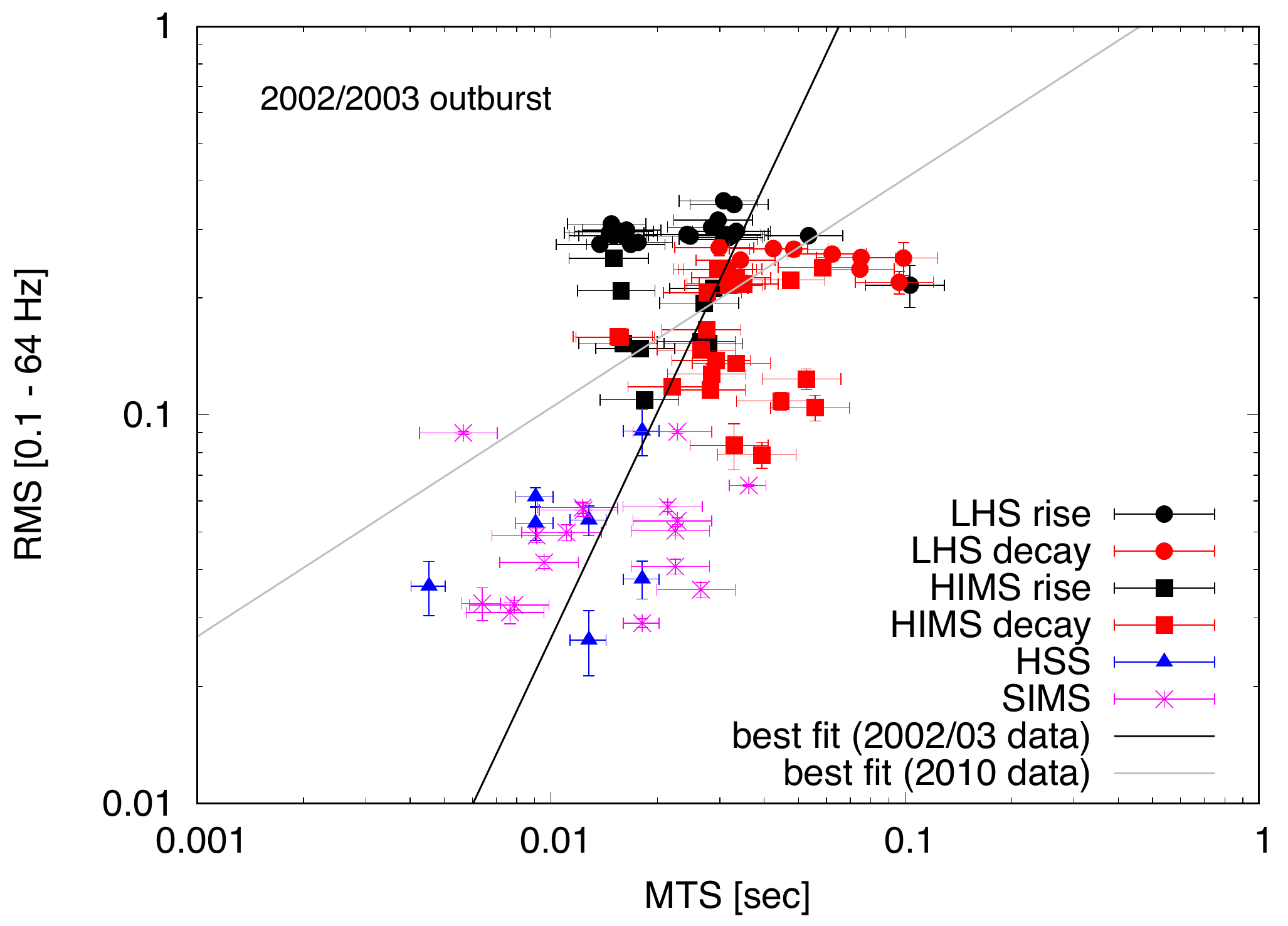}
\includegraphics[scale=0.35, angle = 0]{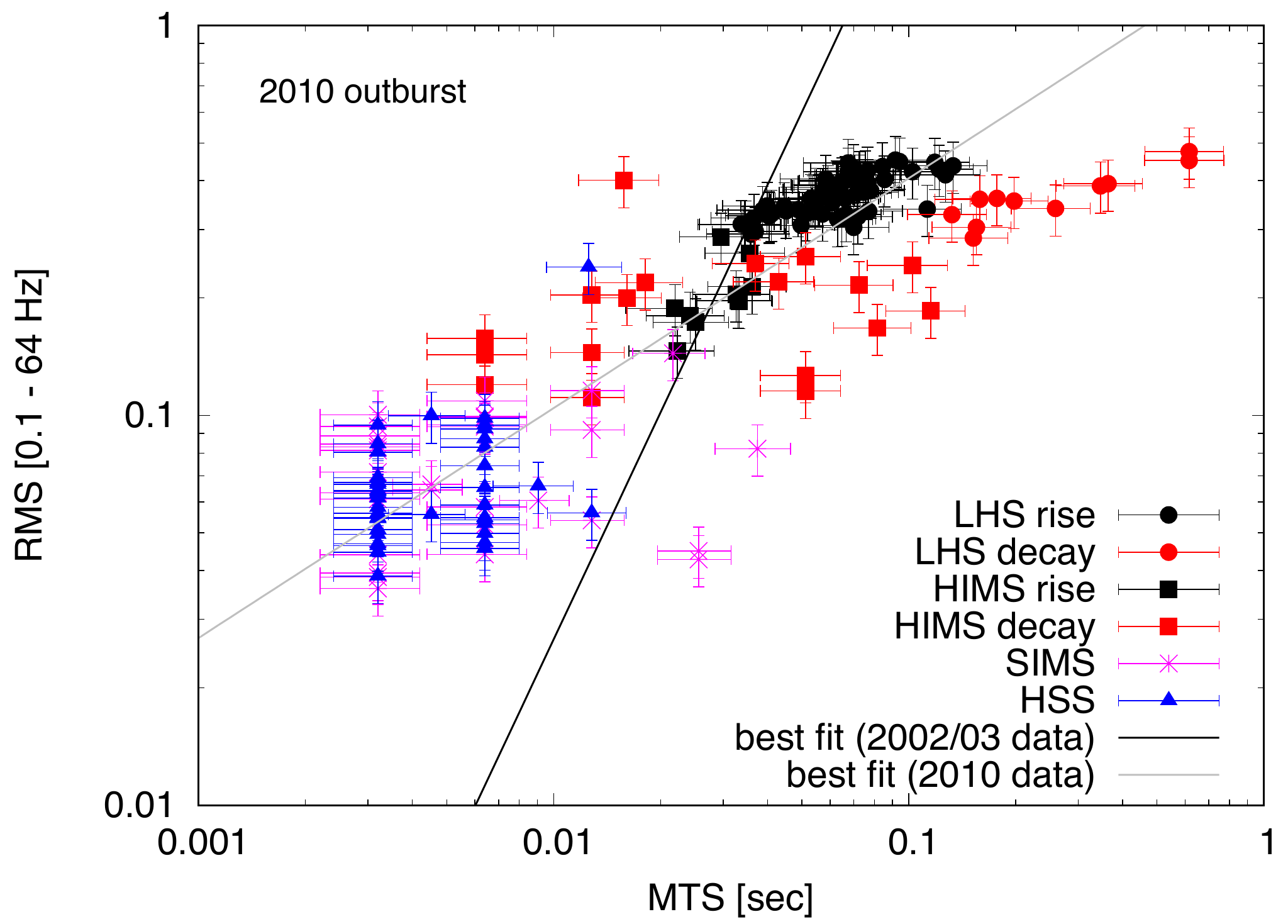}
\caption{Fractional RMS vs. MTS for the 2002-3 and the 2010 outbursts (the upper and lower panels respectively).  The best respective fits are also shown; black line (2002-3 outburst) and the grey line (2010) ouburst. The positive trend is evident for both outbursts, however, the slope of the best-fit line to the 2010 outburst is significantly shallower.}\
    \label{fig7}
\end{figure}
\\
We can put the extracted timescales in perspective by considering the dynamical, viscous and thermal timescales for GX339-4. The dynamical timescale, $t_{dyn}$, is related to the Keplerian frequency $\Omega$, which is given by $\sqrt(GM/r{^3})$. Assuming M = $5M_\odot$ for the BH and r varying from $5R_{g}$ to $25R_{g}$, where $R_{g}$ = $2GM/c{^2}$, we obtain a timescale in the range $\sim$ 5 - 55 ms. The viscous (accretion) timescale, $t_{vis}$, computed as 1/($\alpha(H/r){^2}\Omega$), where we assume the standard disk with the $\alpha$ prescription for the viscosity \citet{1973A&A....24..337S}, with typical values for $\alpha$ and H/r i.e., 0.1 and 0.01 respectively. As expected, the viscous timescale turns out to be significantly longer i.e., extending well over hundreds of seconds compared with dynamical scale. Finally, for completeness, we mention the thermal timescale (measure of the heating and cooling rate of the disk), $t_{thm}$, which as noted by \citet{2016A&A...587A..13L}, is intermediate to the other two scales, and is related to $t_{dyn}$, by 1/$\alpha$. In summary, the $t_{dyn}$ is the shortest timescale and appears to match reasonably well with the extracted MTS but of course with the assumption of a length scale varying between $5R_{g}$ - $25R_{g}$. If proven to be robust, this would suggest strong evidence for the emission region, assumed to be the accretion disk (plus corona), to be truncated in the LHS state (long timescale) and extended (inward toward the BH) in the HSS state (short timescale) over a radial extension that varies by $\sim 20R_{g}$.\\
\\
As noted in the introduction, characteristic timescales can also be extracted directly from PSDs. We have used the PSDs primarily to compute the RMS and have used the time domain to extract the MTS from the lightcurve data. In principle one should be able to make a connection between the two methods. However, in order to accomplish this and uncover the underlying systematics of the apparent RMS-MTS correlation and the source of the dispersion an in-depth investigation is required whereby the roles of physically meaningful parameters that can be linked and/or directly extracted from the PSDs for each state are explored. In the very least, the relevant parameters include the break frequency, a feature that determines the turnover of the PSD at low frequencies, the cutoff frequency, a high frequency feature that represents the signal-noise threshold (most likely related to the MTS), the spectral slope of the PSDs, and finally, a measure of the signal-to-noise ratio which can vary quite significantly as a function of spectral state. Ideally, a study of this kind would involve high statistics data, preferably for several sources, for which PSDs can be constructed for the full range of spectral states. In lieu of high quality PSDs, an alternative approach (the subject of a future study) that could provide useful insight is to simulate PSDs that reflect, realistically as possible, the subtle features observed in the data across all the states. In order to yield reliable results, this approach,  while potentially appealing, demands a balance between the use of simplistic models (such as powerlaws) and arbitrary phenomenological models with a large parameter space that can easily lead to internal correlations and ambiguous interpretation.\\
\\
\begin{figure}
\vspace{-0.1cm}  
\centering      
\includegraphics[scale=0.60, angle = 0]{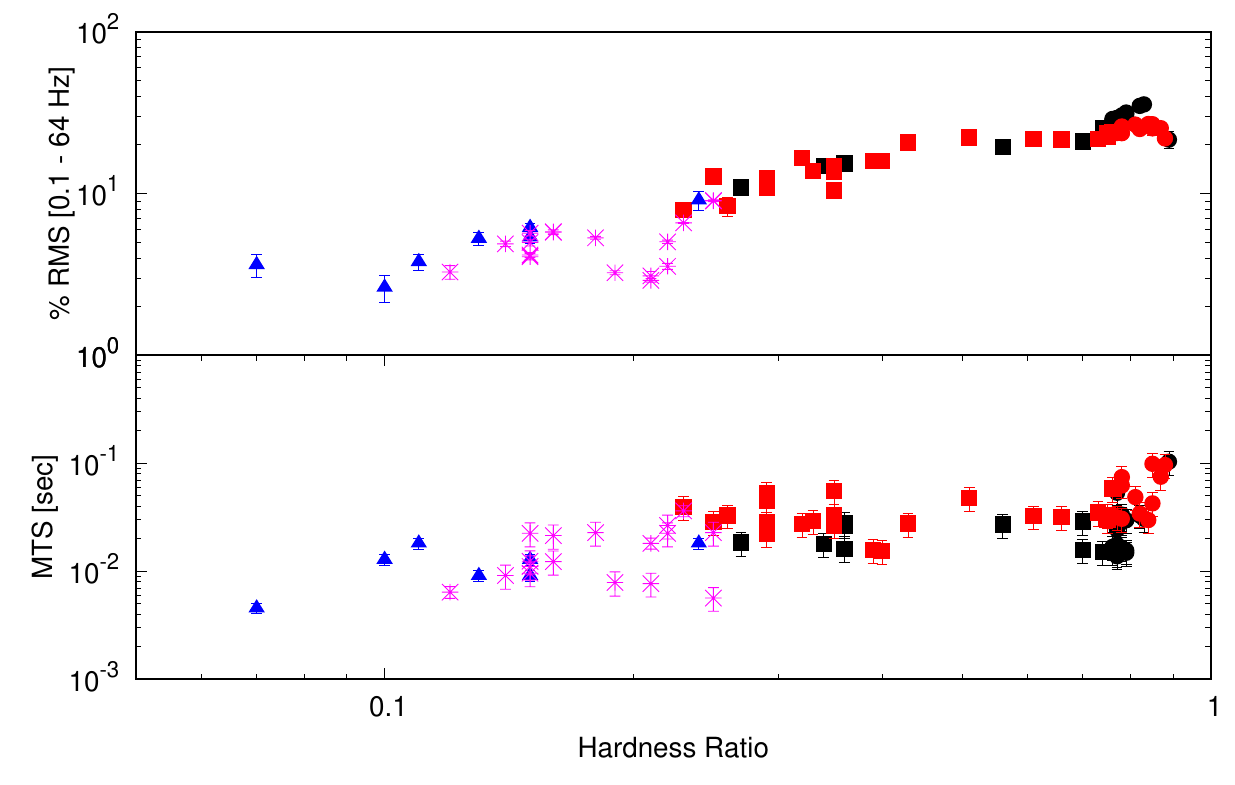}
\caption{Fractional RMS (upper panel) and MTS (lower panel) vs Hardness for the various spectral states of GX339-4 (2002-3 outburst). Symbols and colors as in the previous figures.}
    \label{fig8}
\end{figure}
\begin{figure}
\centering      
\includegraphics[scale=0.60, angle = 0]{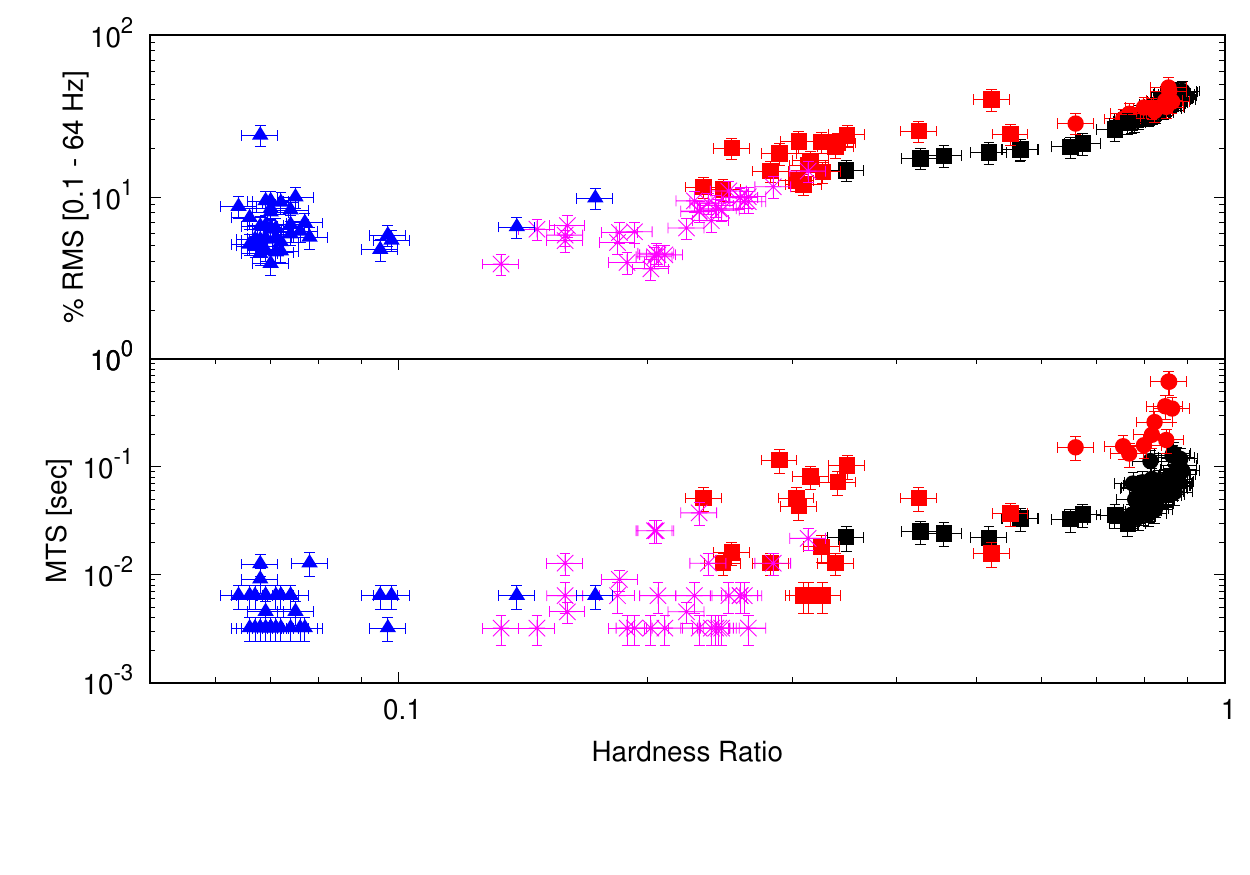}
\caption{Same as in Figure 8 except for the 2010 outburst.}
    \label{fig9}
\end{figure}
Multi-parameter simulations of PSDs notwithstanding, we can nonetheless use the observed power spectral densities to identify and extract the low-frequency turnover of the PSDs i.e., the break frequencies, and determine a "physically" meaningful timescale, and make a comparison with the MTS. In order to accomplish this, we deployed a broken power-law (BPL) spectral model to fit the observed PSDs. Of course, we realize that the BPL is not necessarily an ideal model to describe every possible PSD but it serves our purpose in that we can use it to establish, at least to first order, whether the MTS can be related to a timescale extracted directly from PSDs. The fits to the LHS and HIMS states were relatively straightforward, however, the fits to the PSDs of HSS and SIMS states proved much more challenging because of their irregular features and thus the (few) extracted break frequencies for these data are less reliable and exhibit greater dispersion. The extracted break frequencies are listed in Table 2 (We assigned a 30$\%$ uncertainty to all the break frequencies). Our results are displayed in Figure 10, where we have plotted the break frequencies (as a timescale via a simple inversion) vs. MTS for the 2002-3 and 2010 outbursts. A positive trend is evident as indicated by the best-fit line of slope 2.8 $\pm$ 0.2. Despite the dispersion, the result is strongly suggestive of the MTS being related to a physically meaningful timescale intrinsic to the PSDs. Moreover, the correlation between the two timescales offers a possible explanation for the apparent connection between the RMS and the MTS since the RMS too is a property of the underlying PSD.\\
\begin{figure}
\vspace{-0.1cm}  
\centering      
\includegraphics[scale=0.60, angle = 0]{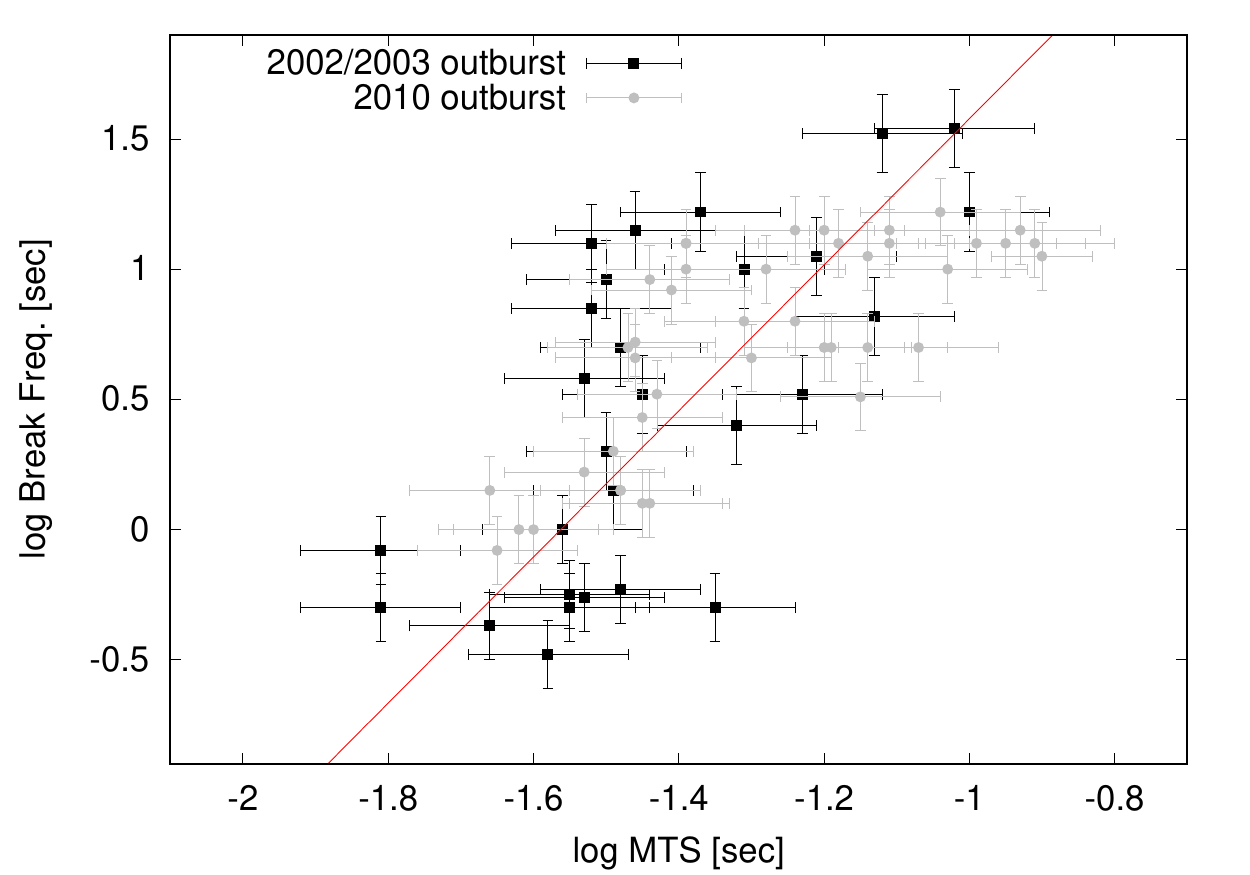}
\caption{Break frequency (sec) vs MTS (sec) for the low-hard and HIMS states of GX339-4 for the 2002-3 and 2010 (grey) outbursts. A positive trend is indicated by the best-fit line with a slope of 2.8 $\pm$ 0.2.}
    \label{fig7}
\end{figure}
\\ 
\begin{figure*}
\centering      
\includegraphics[scale=0.3, angle = 0]{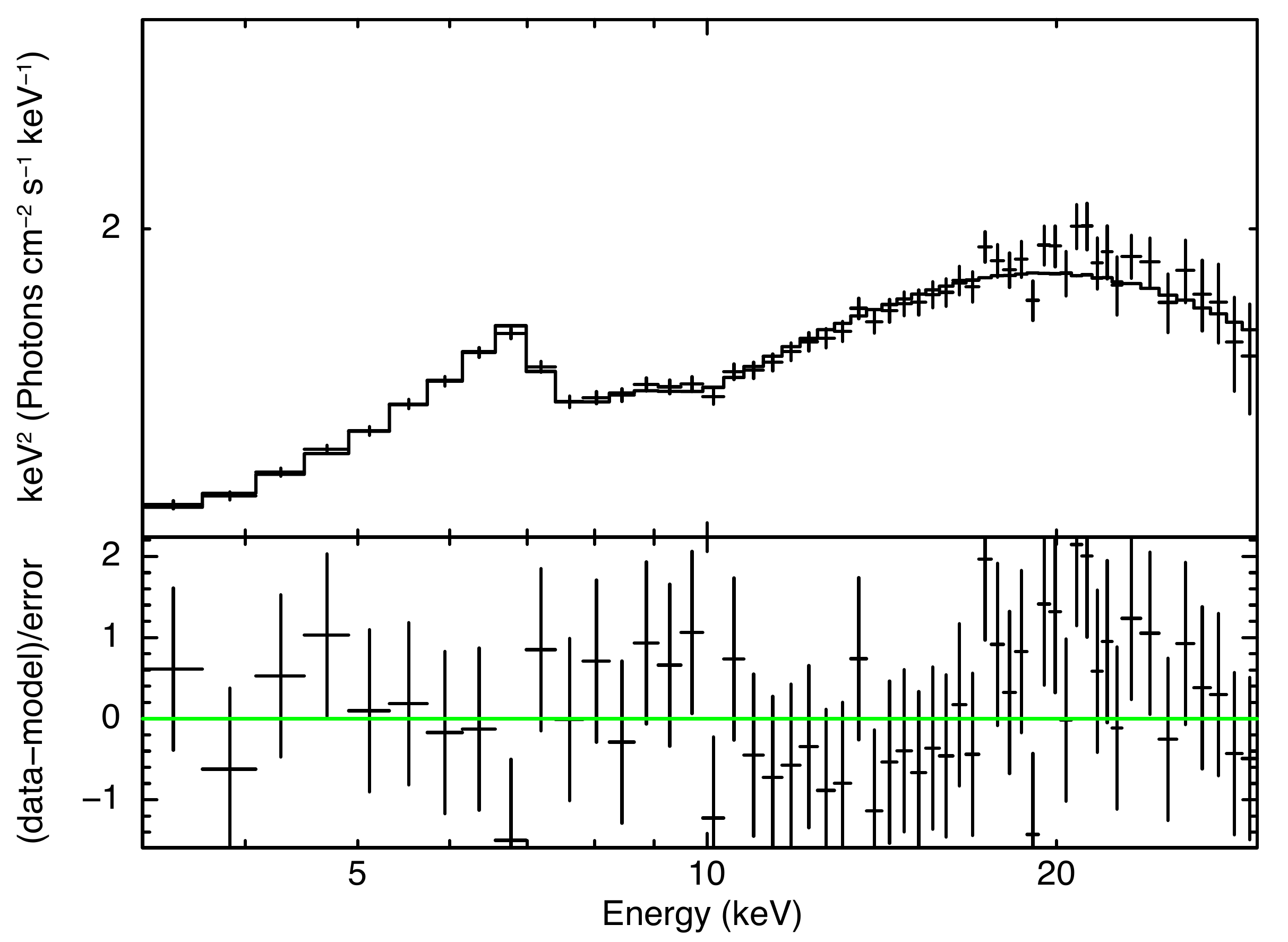}
\includegraphics[scale=0.3, angle = 0]{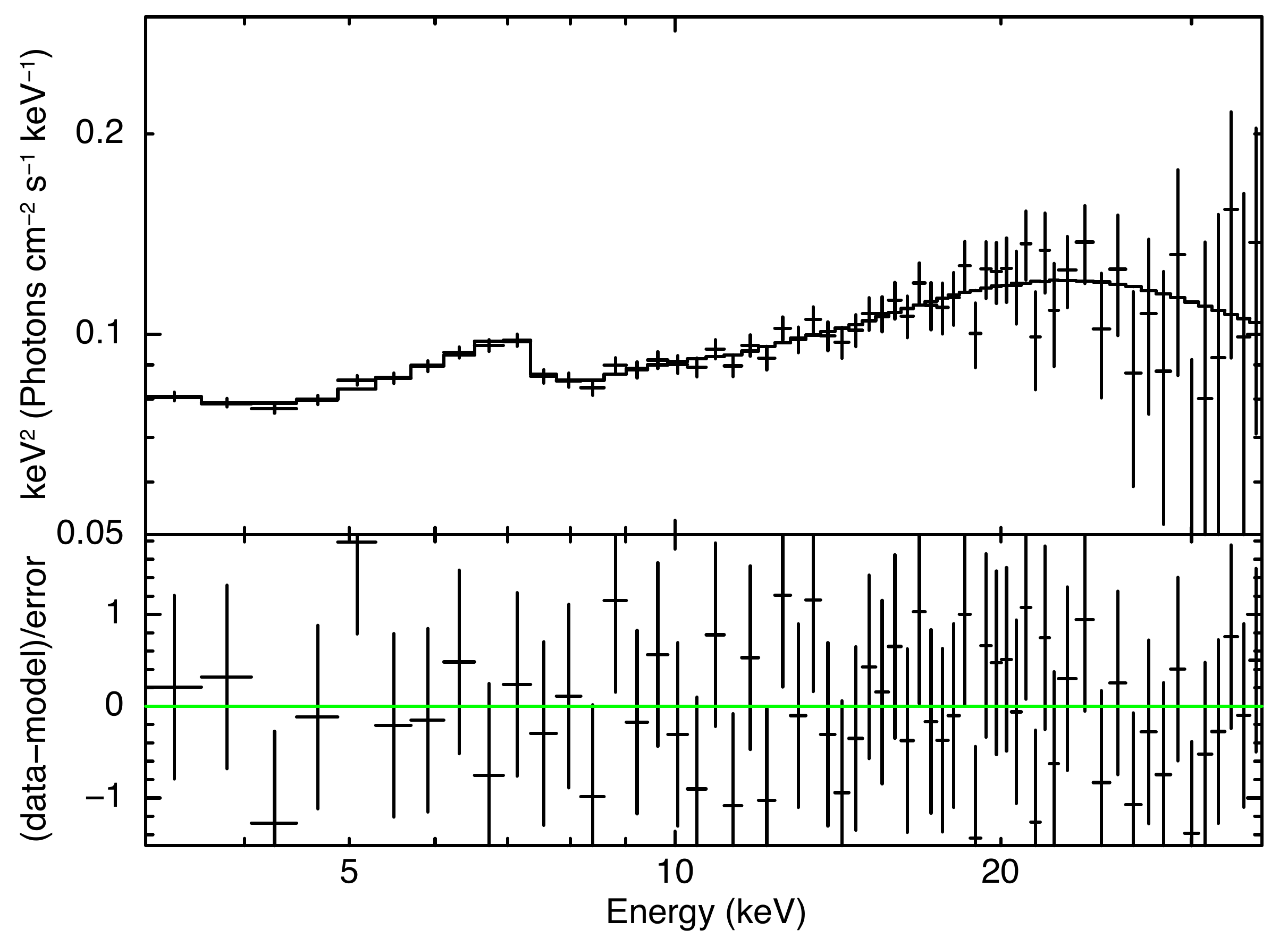}
\includegraphics[scale=0.3, angle = 0]{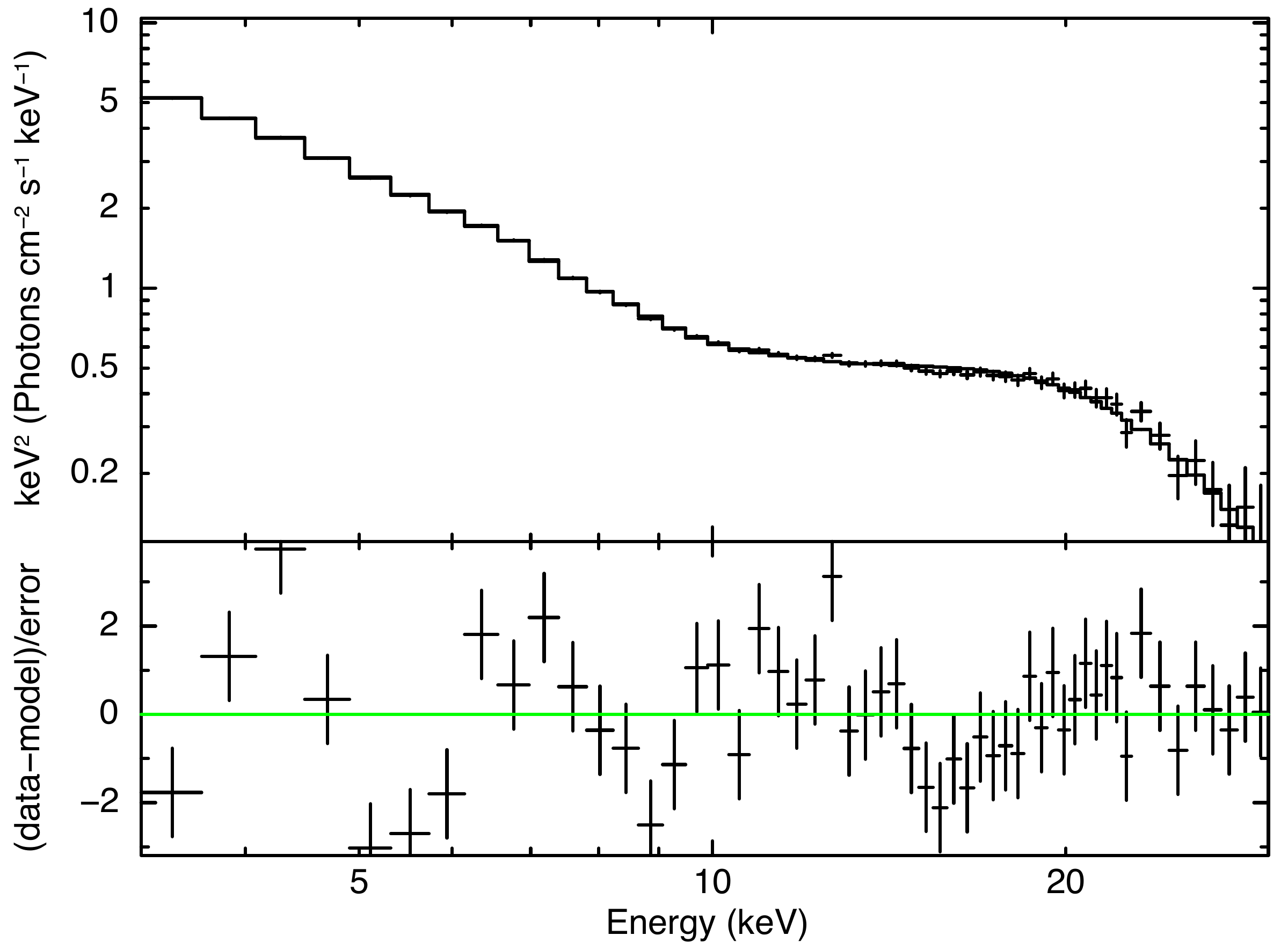}
\includegraphics[scale=0.3, angle = 0]{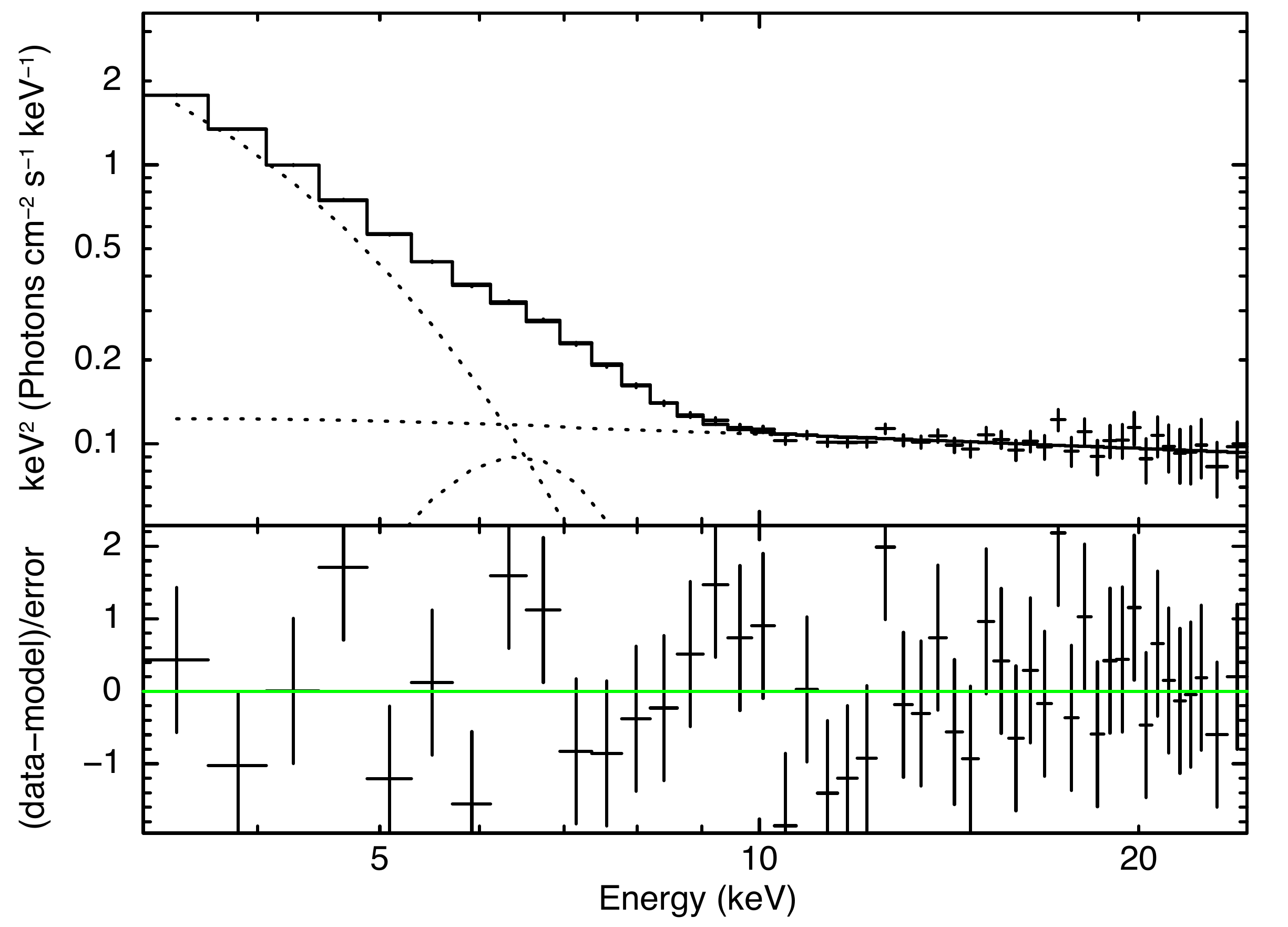}
\caption{Examples of spectral fits with $relxill$, $diskbb$ and $PL$ for the LHS (top left), HIMS (top right), and SIMS (bottom left); the HSS (bottom right) was fitted with $diskbb$, $PL$ and a gaussian.}
\label{fig1}
\end{figure*}
As noted already, the MTS is associated with the high frequency components of the signal and as such it is likely sensitive to the disk/corona density i.e., related to some measure of the interaction time of collisions as the photons undergo multiple scattering in a region where both the viscosity and the radial temperature profile are highly variable. Indeed, this would appear to be consistent with the fact that MTS (on average) is larger for the low-hard and hard-intermediate states, where presumably the emission is dominated by a hot optically thin gas of the corona, as compared to that for the high-soft and soft-intermediate states where the emission is expected to be largely due to a cooler optically thicker gas associated with the disk.\\
\begin{figure}
\centering      
\includegraphics[scale=0.60, angle = 0]{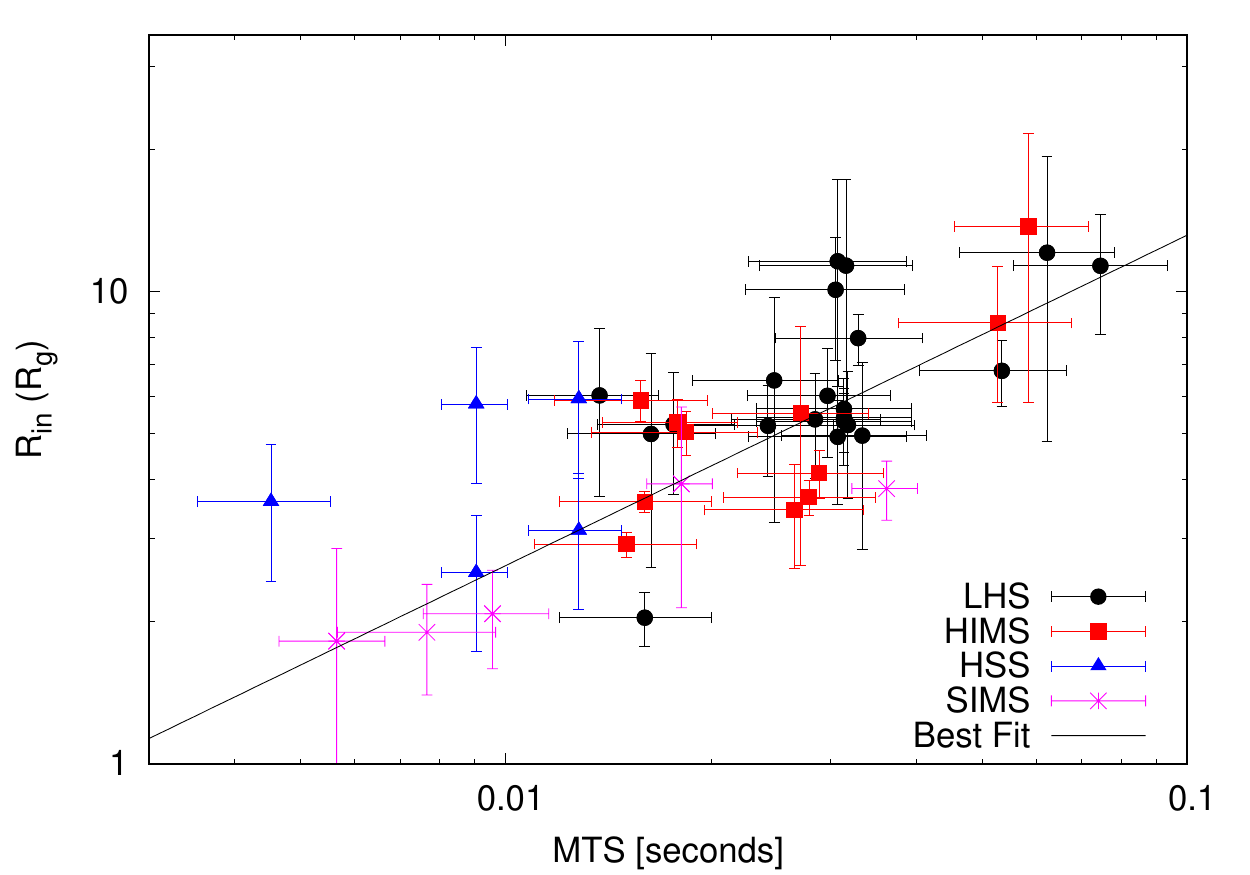}
\caption{Inner disk radius $R_{in}$ vs. MTS for GX339-4 (2002-3 outburst). A positive correlation between $R_{in}$ and MTS is apparent. The solid line, with a slope of 0.70 $\pm$ 0.07, represents a fit to the data. The result is in good agreement with the predicted slope of 2/3 by thermal/viscosity-based models.}
\label{fig1}
\end{figure}
\\
To pursue the possible connection of the MTS with some physical measure of accretion geometry, we followed the procedure laid out by \cite{2015ApJ...813...84G} to track the evolution of the inner disk radius; we included not only the LHS state but also observations for the HSS and the HIMS states in our spectral analysis. We emphasize here that the primary aim is the extraction of a uniform and global trend of the inner disk radius rather than the absolute value or the fine-tuning of model parameters for best statistical outcomes. We are aware that a number of studies, for example \cite{2014MNRAS.442.1767P}, and \cite{2015ApJ...808.122F}, have demonstrated that it is difficult to measure the inner disk radius precisely. These and other studies indicate that the extracted values depend strongly on the assumed geometry and the profile of the accretion disk, particularly the emissivity. With this in mind, we deploy a minimum combination of models for the fits. Of course by taking this approach, we realize that the final spectral fits are unlikley to be optimal and urge caution in the interpretation of the various extracted parameters, particularly the absolute values of $R_{in}$.\\
\\
The models used in fitting the spectral data include the thermal (\textit{diskbb}), phenomenological powerlaw (PL), and the more physically-motivated reflection model, \textit{relxill} \citep{2014ApJ...782...76G}, which not only treats strong gravity but also includes effects of reflection, which are associated with the illumination of the disk by a hot corona in the vicinity of the central object. The known galactic absorption was included explicitly and an intrinsic absorption was left as a free parameter. Mass estimates of the BH in GX339-4 vary in the range 5$\sim$15~M$_\odot$; we assumed a mass of 5M$_\odot$, a distance of 10 kpc and a maximal spin of 0.998 (see \cite{2010MNRAS.403...61D} and references therein). In addition, we fixed the inclination of the disk at 48 $\pm$ 5 degrees (consistent with \cite{2015ApJ...813...84G}) and assumed the disk emissivity to follow the nominal radial dependence i.e., $\propto$ $r^{-3}$. In general, we found the combination of \textit{Tabs*relxill} to provide an acceptable fit for the majority of the spectra in the LHS state. However, contrary to \cite{2015ApJ...813...84G}, who found the Fe abundance to be unusually elevated ($\sim$ 4 $\times$ solar), we found a nominal Fe abundance. Similar to \cite{2015ApJ...813...84G}, we also find the need for a cold absorption component, possibly associated with a distant region of the disk or a disk wind. This effect was well modeled by \textit{relxill} and significantly reduced the residuals in the Compton hump. While the LHS data are well fitted with the \textit{relxill} model, the HSS spectra, on the hand, were better fitted with a combination of \textit{diskbb}, \textit{PL} and a \textit{Gaussian}.  For a number of the HIMS observations the \textit{diskbb} model was included in addition to the \textit{relxill} model in order to improve the fits. For these observations the extracted $R_{in}$ was found to be self-consistent within the two models. Acceptable fits for the SIMS observations required the addition of \textit{Gaussian} absorption. The fit parameters and the models used, are listed in Table 3 along with the observation ids. The typical fits obtained are shown in Figure 11. Finally, we display in Figure 12, the extracted truncation radius $R_{in}$ (in units of $R_{g}$) with the temporal scale MTS and demonstrate a positive correlation between these parameters. The correlation strongly suggests, at least for GX339-4, that the small timescale, is associated with the HSS state (the inner edge of the disk is relatively close to the BH), and that the larger timescale, is associated with the LHS state (the inner edge is further out from the BH). The best-fit line, with a slope of 0.70 $\pm$ 0.07, suggests that the expansion/recession rate is consistent with thermal and/or viscosity driven processes which predict (\citet{2006Natur.444..730M}) a slope of 2/3.  Assuming that $R_{in}$ is a measure of the spatial extent of the emitting region i.e., some combination of a coupled system that includes the inner disk and the surrounding corona, then the expansion/recession rate  associated with the slope provides a measure of the evolution of that region. We note in passing that the correlation between MTS and $R_{in}$ appears to be more robust than the trend between the break frequency and $R_{in}$ reported by \cite{2015A&A...573...120P} who used a very limited number of LHS observations. Our results (Figure 12), on the other hand, include all the dominant spectral states observed in GX339-4. However, we do acknowledge the large uncertainties in our result and the existence of reservations regarding the extraction of $R_{in}$ with the use of simple spectral models and the possible degeneracies among some of the model parameters.  
\section{Summary and Conclusions}
\noindent
In this study, we have analyzed the available archival RXTE data for the 2002-3 and 2010 outbursts of the BH binary GX 339-4. Following the procedures described by \citet{2000MNRAS.318..361N}, \citet{2003A&A...407.1039P} and \citet{2005A&A...440..207B}, we have constructed PSDs for various spectral states of GX339-4 and have extracted the RMS variability by two methods: by fitting the PSDs with a combination of a powerlaw and Lorentzians, and b) by directly integrating the PSDs. We used the technique developed by \cite{2013MNRAS.432..857M}, that employs wavelet basis states, to extract characteristic time scales (MTS) for lightcurves constructed for all the observations across the various spectral states of GX 339-4. In order to probe the connection between the MTS and timescales extractable directly from the observed power spectral densities, we fitted the PSDs with a BPL and extracted break frequencies. Finally, we estimated the inner disk radius, $R_{in}$, for a  number of states by fitting their spectra using simple spectral models available in XSPEC \citep{1996ASPC..101...17A}. We summarize our main findings as follows;\\   
\begin{itemize}
\item We use the extracted MTS to construct an intensity-variability diagram (IVD) for each outburst of GX 339-4. \textit{As far as we are aware, this is the first time such diagrams have been constructed}. The IVD exhibits similar behavior to the q-curve tracks seen in the standard HID plots.\\
\item We link the extracted RMS variabilities to the characteristic time scales for each lightcurve and demonstrate a positive correlation between fractional RMS and MTS, albeit with considerable dispersion. The Pearson correlation coefficient for the 2010 outburst is larger than that for the 2002-3 outburst, indicating a stronger correlation.\\
\item The correlation between the RMS and MTS is not linear thus implying MTS does not follow a simple $\sigma$ (rms)-flux like relation as typically seen in the LHS state. Moreover, the relation is certainly not causal but suggests an independent sensitivity of the two variables to a parameter (or a set of parameters) connected to the underlying emission mechanism(s).\\
\item The RMS-MTS correlation indicates a moderate separation between the LHS and HSS states, with the LHS primarily associated with a larger time scale ($\sim$30 ms) and the HSS with a shorter time scale ($\sim$10 ms). One possible interpretation of these timescales is that the emission region varies in radial extension.\\
\item The break frequencies, extracted by fitting the observed PSDs with a BPL, show a positive trend as function of the MTS. This suggests that the MTS is likely related to a timescale directly attributable to the underlying PSDs.\\
\item We connect the extracted MTS with inner disk radius and demonstrate a positive correlation between them. The slope of the best-fit line is in good agreement with the prediction based on thermal/viscosity-driven accretion processes. As noted elsewhere, we deployed a minimal set of simple models in our spectral fits so the extracted $R_{in}$ values should realistically be considered as best estimates. Nonetheless, in addition to the connection with the break frequencies, the $R_{in}$-MTS trend provides further evidence for the MTS being a physically relatable time scale that has the potential to be useful in gauging the evolution of the accretion geometry in BH binaries.\\
\end{itemize}
\begin{table*}
	\centering
	\caption{RMS and MTS for spectral states in GX 339-4 (2002-2003 outburst). Full version of the table is available online.}
	\label{tab:example_table}
	\begin{tabular}{rrrrrrr} 
	\hline
Obs. Id. & RMS & $\delta$ RMS & $\tau$ & $\delta\tau^{\pm}$  &Break Freq. &PCU2  \\
	\hline
& $\%$ & $\%$ & sec & sec & Hz &cnts/sec  \\
\hline\hline
& & &	{\bf Low-hard state}  & & \\
\hline
60705-01-54-00		&21.6 & 2.7& 0.103&0.026 & & 9  \\
60705-01-55-00		&35.6  &	0.3 & 0.031& 0.008& & 151\\
70109-01-02-00		& 34.8 &	0.1 & 0.033& 0.008 & & 226 \\
70109-01-01-00		&31.7 & 0.1 & 0.030& 0.007& &463 \\
70110-01-01-00			& 31.0  &0.1 & 0.015& 0.004&& 579 \\
70110-01-02-00		& 29.7 &	0.1 & 0.016& 0.004& &600 \\
70109-01-03-00		&30.4 &	0.1 & 0.028&0.007 && 684 \\
70110-01-03-00		& 29.9 &	0.1 & 0.016& 0.004& &766 \\
70110-01-04-00		& 29.4&	0.1 & 0.015& 0.004& &825 \\
40031-03-01-00		& 29.7 &	0.1 &0.033 & 0.008& &835 \\
\hline
\end{tabular}
\end{table*}
\begin{table*}
	\centering
	\caption{RMS and MTS for spectral states in GX 339-4 (2010 outburst). Full version of the table is available online.}
	\label{tab:example_table}
	\begin{tabular}{rrrrrrr} 
	\hline
Obs. Id. & RMS & $\delta$ RMS \tablefootnote{1} & $\tau$ & $\delta\tau^{\pm}$  &Break Freq. &PCU2  \\
	\hline
& $\%$ & $\%$ & sec & sec & Hz & cnts/sec  \\
\hline\hline
& & &	{\bf Low-hard state}  & & &\\
\hline
95409-01-01-00		&45.2 & 6.8& 0.091&0.023& 0.06 & 80  \\
95409-01-02-00		&44.6  &	6.7 & 0.118& 0.030& 0.07&76\\
95409-01-02-01		& 42.2 &	6.3 & 0.102& 0.026&0.08& 91 \\
95409-01-02-02		&44.7 & 6.7 & 0.094& 0.024& 0.10 &104 \\
95409-01-03-02		& 43.5  &6.5 & 0.084& 0.021& 0.20 &112 \\
95409-01-03-03		& 42.8 &	6.4 & 0.122& 0.030& 0.08 &114 \\
95409-01-03-00		&40.3 &	6.0 &0.086&0.021& &106 \\
95409-01-03-04		&40.5 &	6.1 & 0.073& 0.018& 0.09&119 \\
95409-01-03-05		& 40.1&	6.0 & 0.066& 0.017& &136 \\
95409-01-03-01		& 43.7 &	6.5 &0.133 & 0.033& &117 \\
\hline
\end{tabular}\\
$^1$ Given the dispersion observed in the 2002/2003 outburst, we assigned a global 15$\%$ uncertainity for the RMS values calculated for the 2010 outburst.
\end{table*}
\begin{landscape}
\begin{table}
\centering
\caption{Summary of fit parameters for the 2002-2003 outburst of GX 339-4. Full version of the table is available online.}
\begin{tabular}{ccccccccccccc} 
\hline
Obs. Id. & State & Model  & {N$_{H}$} &$R_{in}$ (R$_{g}$) & $\Gamma$  & log xi &A$_{Fe}$ &E$_{c}$ / kT$_{in}$  &R$_{f}$& gauss/gabs&Norm.&$\chi^{2}$/dof \\
\hline
 & & &(10$^{22}$ cm$^{-2}$)&&&&&keV&&E(keV)&&\\
 \hline\hline
40031-03-02-06 & Low-hard & relxill & -- & 6.5$\pm$3.2 & 1.8$\pm$0.1 & 2.9$\pm$0.1 & 0.7$\pm$0.2 & 59.8$\pm$7.2 &  1.3$\pm$0.3&-- &--&72/60\\
70110-01-01-10 & Hard-intermediate & relxill & --& 13.8$\pm$7.9 & 2.1$\pm$0.4 & 2.2$\pm$0.9 & 0.5$\pm$0.1 & 167.0 &  2.0 &-- &--&17/22\\
70110-01-20-00 & Soft-intermediate & relxill*gabs & 1.7$\pm$0.2 & 3.9$\pm$1.8 & 2.6$\pm$0.2 & 4.4$\pm$0.4 & 0.5$\pm$0.2 & 5.3$\pm$0.7 &  0.8 &7.19& --&101/43\\
70110-01-81-00 & High-soft & diskbb+pl+gauss & 0.4 & 5.9$\pm$1.9 & 2.2$\pm$0.1 & --& -- & 0.65$\pm$0.01 &  -- &6.02$\pm$0.14&3652$\pm$180 &46/47\\
\hline
\end{tabular}
\end{table}
\end{landscape}

\section*{Acknowledgements}

This research is supported by the Scientific and Technological Research Council of Turkey (TUBITAK) through the project  number 117F334. In addition, KM acknowledges financial support provided by the Egyptian Cultural and Education Bureau in Washington DC, USA. The Referee's feedback is very much appreciated as it helped to clarify a  number of important issues.

\section*{Data availability}

 The RXTE (Rossi X-ray Timing Explorer) data analyzed in the course of this study are available in HEASARC, the  NASA's Archive of Data on Energetic Phenomena. The data underlying this article are available in the article and in its online supplementary material.






\bsp	
\label{lastpage}
\end{document}